\documentclass{ifacconf}

\usepackage{graphicx}      
\usepackage{natbib}        
\usepackage{color}

\usepackage{amsmath,amssymb,amsfonts}
\usepackage{array}
\usepackage{mathtools}
\usepackage{balance}
\usepackage{setspace}
\newtheorem{proposition}{Proposition} 

\newtheorem{theorem}{Theorem}
\newtheorem{assumption}{{Assumption}}
\newtheorem{lemma}{Lemma}
\newtheorem{definition}{Definition}
\newtheorem{remark}{Remark}

\newtheorem{problem}{Problem}

\newcommand{\td}{\mathrm{d}}     
\newcommand{\tod}{\mathrm{od}}

\newcommand{\mo}[1]{{\color{black} #1}}
\newcommand{\sa}[1]{{\color{black} #1}}

\begin{document}
\begin{frontmatter}

\title{Linear Program-Based Stability Conditions for Nonlinear Autonomous Systems}




\author{Sadredin Hokmi}
and 
\author{Mohammad Khajenejad}

\address{Department of Mechanical Engineering, The University of Tulsa, 
   Tulsa, OK USA (e-mail: \texttt{\{sah3700, mok7673\}@utulsa.edu}).}

\begin{abstract}                
This paper introduces a novel approach to evaluating the asymptotic stability of equilibrium points in both continuous-time (CT) and discrete-time (DT) nonlinear autonomous systems. By utilizing indirect Lyapunov methods and linearizing system dynamics through Jacobian matrices, the methodology replaces traditional semi-definite programming (SDP) techniques with computationally efficient linear programming (LP) conditions. This substitution substantially lowers the computational burden, including time and memory usage, particularly for high-dimensional systems. The stability criteria are developed using matrix transformations and leveraging the structural characteristics of the system, improving scalability. Several examples demonstrated the computational efficiency of the proposed approach compared to the existing SDP-based criteria, particularly for high-dimensional systems.
\end{abstract}


\end{frontmatter}
\section{Introduction}
The stability of nonlinear systems is a cornerstone of control theory, underpinning the safe and reliable operation of a wide range of engineered systems. For instance, in robotics, stability ensures precise motion control and the safe interaction of autonomous systems with their environment, critical for applications such as robotic surgery and industrial automation~[\cite{hogan1985impedance,slotine1987adaptive}]. In power systems, stability analysis is vital for maintaining the balance between generation and demand, preventing cascading failures, and ensuring grid reliability~[\cite{kundur2007power}]. Aerospace engineering relies on stability principles to design robust flight control systems that maintain desired trajectories under disturbances and uncertainties~[\cite{etkin1995dynamics}]. Similarly, in cyber-physical systems (CPS), stability guarantees underpin the synchronization of interconnected systems such as autonomous vehicles and smart grids~[\cite{lee2017introduction}]. Given the diverse and critical nature of these applications, the development of \mo{robust and computationally efficient} methodologies for stability analysis has remained a central focus of control systems research.

Lyapunov-based methods are among the most widely used techniques for stability analysis. These methods rely on constructing a scalar Lyapunov function, which decreases along system trajectories, indicating stability\sa{~[\cite{lyapunov1892general}]}. While these methods are powerful and versatile, their applicability is often limited by the difficulty of finding suitable Lyapunov functions, especially for high-dimensional or complex systems~[\cite{khalil2002nonlinear}]. Moreover, the lack of systematic procedures for constructing such functions poses a significant challenge in practice.

For systems influenced by external inputs, input-to-state stability (ISS) provides a robust framework to analyze stability. ISS extends Lyapunov methods by considering the effects of inputs on system trajectories, making it particularly useful for interconnected systems~[\cite{sontag1989smooth}]. Despite its utility, verifying ISS conditions can be challenging, particularly for systems with nonlinear input dynamics or hybrid behaviors.

Another promising approach is contraction analysis, which focuses on the convergence of system trajectories by identifying contraction metrics~[\cite{tsukamoto2021contraction}]. This method is especially useful for assessing incremental stability across all trajectories rather than solely at equilibrium points. However, the primary limitation lies in finding appropriate contraction metrics, a challenge analogous to that of constructing Lyapunov functions.

In addition to these analytical methods, region of attraction (RoA) analysis has garnered attention for its ability to quantify the set of initial conditions that converge to a desired equilibrium~[\cite{amato2007region}]. While RoA provides valuable insights into practical stability, its application is often hindered by computational complexity, particularly for systems with high-dimensional state spaces~[\cite{vidyasagar2002nonlinear}]. Numerical approaches to estimate RoA are also prone to conservatism and may not generalize well across all system conditions.

Furthermore, modern computational techniques, such as sum of squares (SOS) optimization, have introduced systematic procedures for stability analysis, particularly for polynomial systems. By leveraging semi-definite programming, SOS optimization enables the construction of Lyapunov functions and the verification of stability conditions~[\cite{parrilo2000structured}]. However, these methods are restricted to polynomial or approximable systems and can become computationally infeasible for systems with a large number of states or high-degree polynomials.

On the other hand, stability analysis has been extensively studied for specific classes of nonlinear systems, each with unique characteristics. Positive systems, where state variables remain non-negative for non-negative initial conditions, are analyzed using structural properties, with the works by~[\cite{rantzer2015scalable}] and~[\cite{zhao2019input}] focusing on stability in linear and nonlinear positive systems, respectively. Furthermore, stabilizing static output feedback controllers were parameterized using linear programming (LP), as presented in~[\cite{rami2007controller, rami2011solvability}], while extensions to input-output gain were discussed in the the works by~[\cite{ebihara20111,khajenejad2024optimalcontrolstatic,khajenejad2024optimalcontroldynamic}]. Monotone systems, which preserve the order of state variables, are analyzed using Lyapunov functions and contraction analysis, with~[\cite{belykh2023beyond}] addressing recent advances for systems with non-smooth dynamics. 
However, these methods are often restricted to specific system structures and may not generalize well to a broad class of dynamical systems.

More recently, the work in~[\cite{kawano2021scalable}] examined foundational results concerning the stability and dissipativity of positive linear systems. It demonstrates how scalable conditions derived from linear programs (LP) can be generalized to encompass cone-invariant positive linear systems. In addition, the research in~[\cite{antal2024computationally}] introduced a sampling-based method that leverages quadratic Lyapunov function parametrization and $\ell_1$-norm optimization to efficiently estimate the domain of attraction (DoA) of nonlinear systems using linear programming, addressing speed and scalability challenges. As demonstrated in~[\cite{ahmadi2008non}] and~[\cite{ahmadi2008non}], the conservativeness of quadratic functions can be mitigated by using non-monotonic Lyapunov functions, where the dynamics are also incorporated within the parametrization.

\sa{Furthermore, in~[\cite{andersen2023lyapunov}], a comparative study of SDP and LP approaches for stability analysis of switched linear systems demonstrated that LP, combined with matrix preconditioning and a visualization tool (Angle Analysis), significantly improves computational efficiency and scalability. The LP method outperforms SDP in solving LMIs and in identifying piecewise linear Control Lyapunov Functions, particularly in high-dimensional systems. Inspired by this, we aim to propose an LP-based method for analyzing stability of nonlinear systems.}

\textbf{\emph{Contributions}.} Despite the aforementioned advancements, existing stability methods often face common drawbacks, including scalability to large-scale systems and computational cost. To bridge this gap, this paper provides computationally efficient linear program (LP)-based conditions for the asymptotic stability of equilibrium points in nonlinear continuous-time (CT) and discrete-time (DT) autonomous systems. By leveraging indirect Lyapunov methods and linear approximations via Jacobian matrices, the proposed approach replaces standard semi-definite program (SDP)-based criteria with LP-based tests, significantly reducing computational complexity for high-dimensional systems. The stability conditions are derived through matrix transformations and structural properties, ensuring accuracy and scalability. These results offer a practical alternative to traditional stability analysis methods, especially in large-scale applications.

\section{Preliminaries}

\textbf{\emph{Notation}.} $\mathbb{N}_n$, $\mathbb{N}$, $\mathbb{R}^{n \times p}$, $\mathbb{R}^n$, and $\mathbb{R}^n_{>0}$ represent, respectively, the natural numbers up to $n$, the set of natural numbers, matrices of size $n$ by $p$, $n$-dimensional Euclidean space, and positive vectors of dimension $n$. For a vector $v \in \mathbb{R}^n$, the $p$-norm is defined as $\|v\|_{p} \triangleq \left(\sum_{i=1}^n {|v_i|^p}\right)^{\frac{1}{p}}$. 
For a matrix $M \in \mathbb{R}^{n \times p}$, the $i$-th row and $j$-th column entry is denoted by $M_{ij}$. 
We define $M^{\oplus} \triangleq \max(M, \mathbf{0}_{n \times p})$, {$M^{\ominus} \triangleq M^{\oplus} - M$, and $|M| \triangleq M^{\oplus} + M^{\ominus}$}, which represents the element-wise absolute value of $M$. Furthermore, $M^\text{d}$ denotes the diagonal matrix containing only the diagonal entries of the square matrix $M \in \mathbb{R}^{n \times n}$, while $M^\text{od} \triangleq M - M^\text{d}$ contains only the off-diagonal elements. The matrix $M^{\text{m}} \triangleq M^\text{d} + |M^\text{od}|$ is known as the ``Metzlerized'' matrix\footnote{A Metzler matrix is a square matrix with non-negative off-diagonal elements.}. Additionally, we use $M \succ 0$ and $M \prec 0$ (or $M \succeq 0$ and $M \preceq 0$) to indicate that $M$ is positive definite and negative definite (or positive semi-definite and negative semi-definite), respectively. Lastly, all vector and matrix inequalities are  element-wise inequalities, and the zero and ones matrices of size $n \times p$ are denoted by $\mathbf{0}_{n \times p}$ and $\mathbf{1}_{n \times p}$, respectively, while $\mathbf{0}_n$ denotes a zero vector in $\mathbb{R}^n$, while $I_n$ is the identity matrix in $\mathbb{R}^{n \times n}$. 

Next, we recall the standard linear matrix inequalities (LMI)s for the stability of linear systems, which we refer to them multiple times throughout the paper.
\begin{proposition}~[\cite{boyd1994linear,bernstein2018scalar}]\label{prop:SDP_stable}
The linear discrete-time (DT) or continuous-
time (CT) system $\mathcal{G}:x^+_t=Ax$, where $x^+_t=x_{t+1}$ if $\mathcal{G}$ is DT and $x^+_t=\dot x_t$ if $\mathcal{G}$ is CT, is asymptotically stable if and only if there exists a positive definite matrix $P \in \mathbb{R}^n$ that solves the following linear matrix inequalities (LMI):
\begin{align}\label{eq:SDP_stable}
\begin{cases}A^\top P+PA \prec 0, \quad \hfill \text{\emph{if}} \ \mathcal{G} \ \text{\emph{is CT}}, \\
APA^\top-P \prec 0, \quad \hfill \text{\emph{if}} \ \mathcal{G} \ \text{\emph{is DT}}.
\end{cases}
\end{align}
\end{proposition}
\sa{Note that checking eigenvalues is basically equivalent to solving the LMIs. Also, the LP stability feasibility conditions are thoroughly developed in Section~\ref{sec:LP_stability}. Furthermore, reviewing~[\cite{rantzer2015scalable}] can be beneficial for further insight into related LP-based stability analysis methods.}

\section{Problem Formulation} \label{sec:problem_formulation}

\noindent\textbf{\emph{System dynamics.}} In this work, we consider autonomous discrete-time (DT) and continuous-time (CT) nonlinear systems, defined as:
\begin{align}\label{eq:system}
x_t^+=f(x_t),
\end{align}
for every $t \in \mathbb{T}$, where {if \eqref{eq:system} is a DT system, then $x_t^+ = x_{t+1}$ and ${\mathbb{T}} = \{0\} \cup \mathbb{N}$; if \eqref{eq:system} is a CT system, then $x_t^+ = \dot{x}_t$ and ${\mathbb{T}} = \mathbb{R}_{\ge 0}$}.
Here, ${f}:\mathbb{R}^n  \to \mathbb{R}^n$ is the nonlinear state vector field. Moreover, we make the following assumption:
\begin{assumption}\label{ass:diff}
The vector field $f$ is a differentiable function for all degrees of differentiation, i.e., $f \in C^{\infty}$.
\end{assumption}
Furthermore, we assume that $x_e$ is a known \emph{equilibrium point} of the system \eqref{eq:system}, i.e., setting $x_0=x_e$ implies $x_t=x_e, \forall t \in \mathbb{T}$ (a degenerate trajectory of the system):
\begin{align}\label{eq:equilibrium}
\begin{cases} f(x_e)=0, \quad \hfill \text{CT case}, \\ f(x_e)=x_e, \quad \hfill \text{DT case}.
\end{cases}
\end{align}
\begin{definition}[Asymptotic Stability]\label{defn:as_stability}
The equilibrium point $x_e$ is asymptotically stable (AS) if:
\begin{enumerate}
\item it is stable, i.e., $\forall \epsilon >0, \exists \ \delta(\epsilon)>0 \ \text{s.t.} \ \|x_0-x_e\|_p < \delta(\epsilon) \Rightarrow  \|x_t-x_e\|_p < \epsilon, \ \forall t \in \mathbb{T},$ and
\item $\exists \ \delta_a >0 \ \text{s.t.} \ \|x_0-x_e\|_p < \delta_a \Rightarrow \lim\limits_{t \to \infty} \|x_t-x_e\|_p=0.$
\end{enumerate}
\end{definition}
Our objective is to provide alternative sufficient conditions for asymptotic stability of equilibrium points in nonlinear systems that are computationally efficient (in the sense of time and memoty usage), especially for high-dimensional systems. This can be formally stated as follows.
\vspace{.3cm}
\begin{problem}\label{prob:LP_stability}
Given the nonlinear dynamics \eqref{eq:system}, provide sufficient linear program-based conditions for the asymptotic stability of any given equilibrium point $x_e$.
\end{problem}

\section{LP Stability Conditions for Nonlinear Systems}\label{sec:LP_stability}
To address Problem \ref{prob:LP_stability}, inspired by \emph{indirect Lyapunov Methods}~[\cite[Chapter 4]{khalil2002nonlinear}] \mo{(as well as many other tractable stability methods for nonlinear systems),} we first compute a linear approximation of \eqref{eq:system}.
\vspace{.3cm}
\begin{definition}[Linear Approximation]\label{defn:lienarization}
Consider a nonlinear system in the form of \eqref{eq:system}, with equilibrium point $x_e$, that is $f(x_e)=0$ in the CT case and $f(x_e)=x_e$ in the DT case. If $f \in C^{\infty}$, then Taylor expansion provides:
\begin{align*}
x^+_t&=f(x_t)=f(x_e)+\frac{df}{dx}\big{|}_{x_e}(x_t-x_e)+h(x_t-x_e)\\
&=\begin{cases}J_f(x_e)(x_t-x_e)+h(x_t-x_e),\quad \hfill \text{\emph{CT case}},\\
x_e+J_f(x_e)(x_t-x_e)+h(x_t-x_e),\quad \hfill \text{\emph{DT case}},
\end{cases}
\end{align*}
where $h(x-x_e)$ collects the (infinite) terms of degree higher than one and $J_f(x_e)$ is the Jacobian matrix of $f$ with respect to $x$, computed at $x_e$. In the new coordinates $$\xi_t\triangleq x_t-x_e \Rightarrow \xi^+_t=\begin{cases} x^+_t \quad \hfill \text{\emph{CT case}},\\
x^+_t-x_e \quad \hfill \text{\emph{DT case}},\end{cases}$$ the dynamics is described by:
$$\xi^+_t=J_f(x_e)\xi+h(\xi).$$ In the vicinity of the equilibrium point $x_e$, higher order terms may be neglected, hence we obtain the following linear approximation for the original system \eqref{eq:system}:
\begin{align}\label{eq:linear_approximation}
\xi^+_t=A\xi_t, \ \text{where} \ A=J_f(x_e).
\end{align}
\end{definition}
Driven by the \emph{indirect Lyapunov Criterion} from which  asymptotic stability of a nonlinear autonomous system at its equilibrium point can be studied based on the stability of the corresponding linear approximation system, one can leverage the semi-definite program (SDP)-based criteria in Proposition \ref{prop:SDP_stable} to investigate the stability of the original system \eqref{eq:system}. Alternatively, we aim to provide linear program (LP)-based conditions, that are known to be computationally more efficient compared to the SDPs especially for high-dimensional systems. To do so, inspired by the existing LP-based criteria for the stability of positive/cooperative systems, we construct a corresponding $2n$-dimensional positive/cooperative linear system to the original approximation system \eqref{eq:linear_approximation}, whose stability implies the stability of \eqref{eq:linear_approximation}. This is done via the following Lemma.  
\begin{lemma}[LP Stability of Linear Systems]\label{lem:LP_stable_linear}
The approximation system in \eqref{eq:linear_approximation} is stable, i.e. $A$ is Schur stable in the DT case and is Hurwitz stable in the CT case, if the following linear program is feasible:
\begin{align}\label{eq:LP_stable}
\exists p \in \mathbb{R}^{2n}_{>0}, \ \text{s.t.} \ \hat{A}p  <  \sigma,
\end{align}
where
\begin{align}\label{eq:A_hat}
\hat{A} \triangleq \begin{bmatrix} A^{\uparrow} & A^{\downarrow} \\ A^{\downarrow} & A^{\uparrow} \end{bmatrix}.
\end{align}
Moreover,
\begin{itemize}
\item If \eqref{eq:linear_approximation} is a DT system, then 
\begin{align*}
A^{\uparrow}=A^\oplus \triangleq \max(A,0), \ 
A^{\downarrow}=A^{\ominus} \triangleq A^\oplus -A,\
\sigma=p.
\end{align*}
\item If \eqref{eq:linear_approximation} is a CT system, then 
\begin{align*}
\hspace{-.3cm}A^{\uparrow}=A^{\td}+A^{\tod,\oplus}, \qquad \quad 
A^{\downarrow}=A^{\tod,\ominus},   \qquad
\sigma=\mathbf{0}_{2n}.
\end{align*}
\end{itemize}
Finally $A^{\td}_{ij}=A_{ij}$ if $i=j$ and $A^{\td}_{ij}=0$ otherwise, while $A^{\tod}=A-A^{\td}$. 
\end{lemma}
\begin{pf}
First, note that by construction, $A^{\downarrow}$ is always a non-negative matrix, while $A^{\uparrow}$ is non-negative in the DT case, and is a Metzler matrix in the CT case. Consequently, the $2n$ dimensional $\hat{A}$ is a positive matrix in the DT case and a Metzler matrix in the CT case. Then, by applying~[\cite[Propositions 2 \& 1]{rantzer2015scalable}] matrix $\hat{A}$ is Hurwitz stable in the CT case, and is Schur stable in the DT case, respectively, if and only if the LP in \eqref{eq:LP_stable} is feasible. In other words, \eqref{eq:LP_stable} implies the stability of $\hat{A}$. What remains is to show this is sufficient for the stability of \eqref{eq:linear_approximation}, i.e., the matrix $A$. To do this, we consider applying a similarity transformation by pre- and post-multiplication of $\hat A$ by $T=T^{-1}=\begin{bmatrix} I_n & I_n \\ 0 & -I_n \end{bmatrix}$, which returns:
$T\hat{A}T^{-1}=\tilde{A}=\begin{bmatrix} |A| & 0 \\ -A^{\downarrow} & A \end{bmatrix}$, where $|A|\triangleq A^{\uparrow}+A^{\downarrow}$. The resulting matrix $\tilde{A}$ is a block lower triangular matrix, hence its set of eigenvalues  is the union of the sets of eigenvalues of its main diagonal block matrices, i.e, $|A|$ and $A$. So, if $\hat A$ (and consequently $\tilde A$) be stable, then $A$ must be stable as well.
\end{pf}
\sa
{\begin{remark}[Neglecting Higher Order Terms]\
In our analysis, we construct $\hat{A}$ to be Metzler in the continuous-time (CT) case and positive in the discrete-time (DT) case. This, along with the LMIs we propose that ensure Hurwitz stability in the CT case and Schur stability in the DT case, guarantees that all eigenvalues have strictly negative real parts, thus eliminating the ambiguity typically associated with marginal stability. Consequently, we can neglect the higher-order terms locally without loss of generality, which justifies the use of the indirect method for local stability analysis. This assumption is standard in the analysis of monotone and positive systems, as also supported in~[\cite[Proposition 1]{rantzer2015scalable}].
\end{remark}
}
\begin{remark}
It is worth noting that applying a similarity transformations will transform the $\hat{A}$ matrix in \eqref{eq:A_hat} into $\begin{bmatrix} A^{\uparrow} & -A^{\downarrow} \\ -A^{\downarrow} & A^{\uparrow} \end{bmatrix}$, which is the state matrix of a $2n$-dimensional so-called tight linear \textbf{embedding system}~[\cite{khajenejad2023tight,khajenejad2024optimalbook}] for \eqref{eq:linear_approximation}. As a known fact, the stability of such corresponding embedding system implies the stability of the original system~[\cite{smith2008global}]. Therefor, the result in Lemma \ref{lem:LP_stable_linear} can be considered as an alternative interpretation of this fact. 
\end{remark}
\begin{remark}
In the special case that the system in \eqref{eq:linear_approximation} is positive/cooperative, the LP conditions in \eqref{eq:LP_stable} are necessary and sufficient, and coincide with the LP conditions for the stability of positive systems in~[\cite[Propositions 1 and 2]{rantzer2015scalable}].
\end{remark}
We conclude this section by leveraging Lemma \ref{lem:LP_stable_linear} to provide LP-based stability conditions for equilibrium points of nonlinear autonomous systems via the following theorem.
\begin{theorem}\label{thm:nonlinear_LP_stability}
Consider the nonlinear autonomous system in \eqref{eq:system} and suppose Assumption \ref{ass:diff} holds. Let $x_e$ be a given equilibrium point and $A=J_f(x_e)$. Then, $x_e$ is asymptotically stable if the LP in \eqref{eq:LP_stable} holds with $\hat{A} \triangleq \begin{bmatrix} A^{\uparrow} & A^{\downarrow} \\ A^{\downarrow} & A^{\uparrow} \end{bmatrix}$, where $A^{\uparrow}$ and $A^{\downarrow}$ are defined in Lemma \ref{lem:LP_stable_linear}.
\end{theorem}
\begin{pf}
The proof follows from Lemma \ref{lem:LP_stable_linear} and the fact that the equilibrium point is AS if the linearized approximation system is stable~[\cite[Theorem 4.7]{khalil2002nonlinear}]. 
\end{pf}
\vspace{-.2cm}
\mo{
\begin{rem}
The matrix variable $P$ in the LMIs in \eqref{eq:SDP_stable} has $n(n+1)/2$ independent entries, and the LMI constraint involves matrix inequalities of size $n \times n$. When solved using an interior-point method, the worst-case arithmetic complexity is typically on the order of: $T_{\text{SDP}} = \mathcal{O}(n^6)$,
since the per-iteration cost involves forming and factorizing large Hessian or Schur complement matrices related to the $n^2$ variables. On the other hand, the linear program in \eqref{eq:LP_stable}--\eqref{eq:A_hat} has $2n$ decision variables and a number of inequality constraints that scale linearly with $n$. For standard interior-point methods applied to LPs, the per-iteration complexity is roughly $T_{\text{LP}} = \mathcal{O}((2n)^3) = \mathcal{O}(n^3)$,
exhibiting lower computational complexity of the LP method compared to the SDP one.
\end{rem}
}
\vspace{-.3cm}
\section{Illustrative Examples}
To illustrate the effectiveness of our approach, we considered six nonlinear dynamical CT systems and their DT versions through discretizations. The systems have different dimensions, i.e., number of states. To investigate the stability of each of the example systems, we applied the well-known SDP-based criteria in Proposition  \ref{prop:SDP_stable} (will be called SDP method/test), as well as the proposed LP-based criteria summarized in Theorem \ref{thm:nonlinear_LP_stability} (will be called LP method/test), for both CT and DT cases. For each system, the corresponding time and memory usage for each of the SDP and LP tests to be executed were computed and compared in Figure \ref{fig:DT_time} and Table \ref{tab:dt_case}, for both DT and CT cases. The LMIs were solved using YALMIP~[\cite{Lofberg2004}] and MOSEK~[\cite{mosek}], and all the simulations were conducted on a PC with 8 GB of RAM and an Intel Core i5-7200U CPU (4 cores, 3.1 GHz). 
The considered examples are as follows.
\subsection{Example 1: CSTR System}
\begin{figure}[htbp]
\begin{center}
\includegraphics[width=4.2cm]{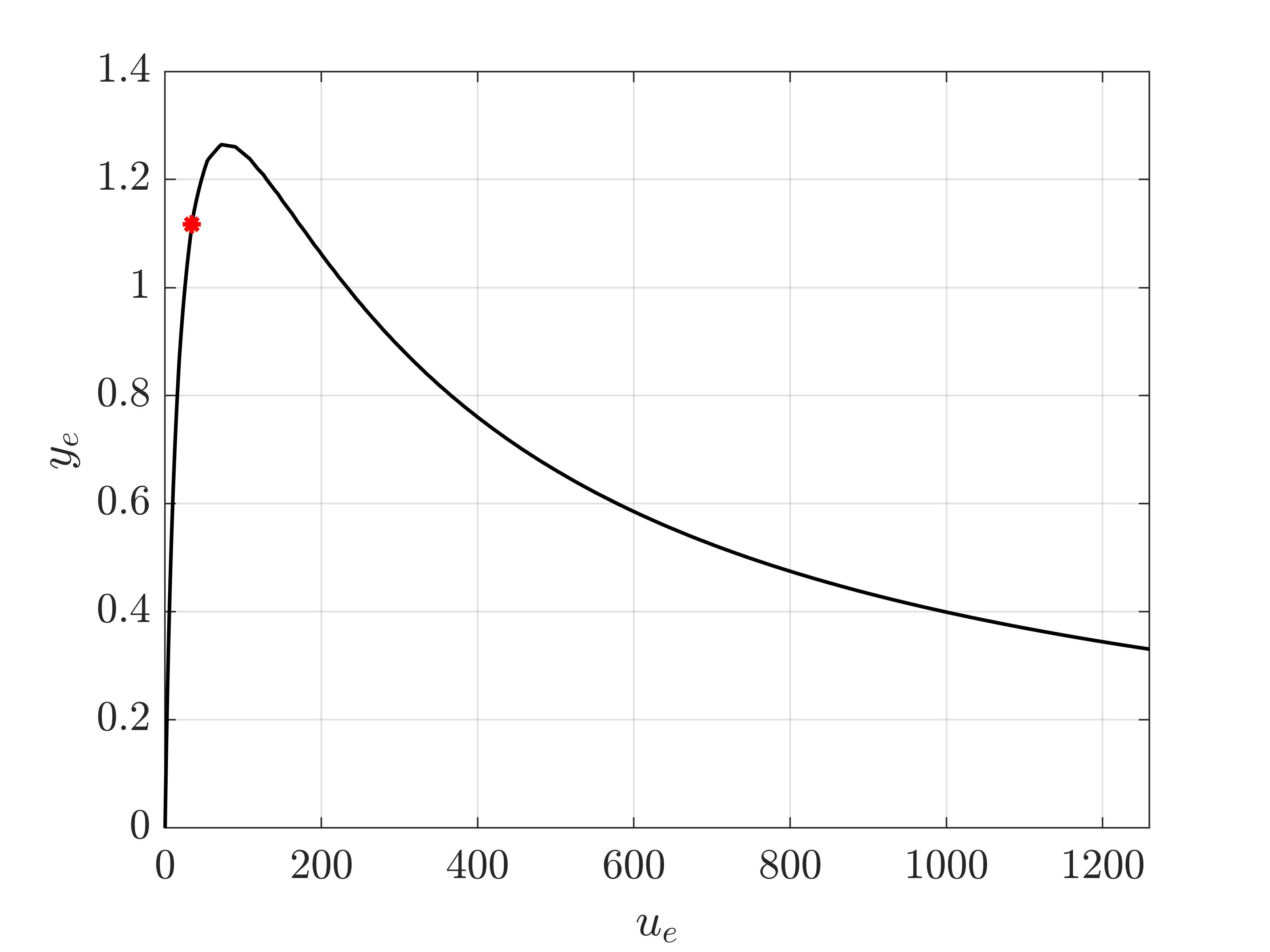}    
\vspace{-.2cm}
\caption{\small Steady-state input-output diagram of the CSTR system.\label{fig:CSTR}}
\end{center}
\end{figure}

As the fist example, we considered a two-dimensional continuous stirred tank reactor (CSTR) system, as described in~[\cite{zhang2023output}]:
\begin{equation*}
\left\{
\begin{aligned}
\dot{x}_1 &= -50x_1 - 10x_1^2 + (10 - x_1)u, \\
\dot{x}_2 &= 50x_1 - 100x_2 - x_2^2.
\end{aligned}
\right.
\end{equation*}
The system output is also considered to be \( x_2 \) (\( y = x_2 \)). Figure \ref{fig:CSTR} illustrates the steady-state input-output diagram of the system, with the operating point highlighted in red.
Based on Figure \ref{fig:CSTR}, the operating point is $(u^*, x_2^*)=(34.288, 1.117)$. Accordingly, the operating (equilibrium) point for $x_1$ is also equal to 3, and hence, $(x_1^*, x_2^*)=(3, 1.117)$. The matrices $A_c$ and $A_d$, representing the CT and DT state matrices, are as follows (with sampling time $T_s = 0.01 s$):
\[
A_c = 
\begin{bmatrix}
-144.286 & 0 \\
50 & -134.286
\end{bmatrix}, \quad
A_d = 
\begin{bmatrix}
0.2363 & 0 \\
0.1242 & 0.2611
\end{bmatrix}.
\]
In the CT case, the execution time for the LP and SDP tests were {$t^{l}_{c}=0.448 s,t^{s}_{c}=0.540 s$}, respectively, while in the DT case, we obtained {$t^{l}_{d}=0.353 s,t^{s}_{d}=0.463 s$}. As expected, for this low dimensional system, the differences between the execution times are negligible.
\subsection{Example 2: Coupled Oscillator}
As for the second example, we considered a five-dimensional coupled oscillator system with damping and nonlinear interactions~[\cite{strogatz2018nonlinear}], with the following dynamics:
\begin{equation*}
\left\{
\begin{aligned}
\dot{x}_1 &= x_2, \ 
\dot{x}_2 = -\alpha_1 x_1 - \beta_1 x_2 + \gamma_1 x_1^3, \
\dot{x}_3 = x_4, \\
\dot{x}_4 &= -\alpha_2 x_3 - \beta_2 x_4 + \gamma_2 x_3^3, \
\dot{x}_5 = -\delta x_1 + \varepsilon (x_1^2 + x_3^2),
\end{aligned}
\right.
\end{equation*}
where \( x_1, x_3 \) represent the positions of two oscillators; \( x_2, x_4 \) represent the velocities of the two oscillators; \( x_5 \) denotes the coupled energy-related state; \( \alpha_1, \alpha_2 > 0 \) are the restoring force constants; \( \beta_1, \beta_2 > 0 \) are the damping constants; \( \gamma_1, \gamma_2 > 0 \) represent the nonlinear terms; and \( \delta > 0, \varepsilon > 0 \) are the coupling parameters.
The equilibrium point is   
\( x_e = \mathbf{0}_4 \).
The sampling time and parameters are selected as \( T_s = 0.1 s, \, \alpha_1 = 2, \, \beta_1 = 3, \, \alpha_2 = 1.5, \, \beta_2 = 2.5, \, \text{and} \, \delta = -4 \). The matrices \( A_c \) and \( A_d \), i.e., the CT and DT state matrices are:

\vspace{-.3cm}
{\small
\begin{align*}
A_c\hspace{-.1cm}=\hspace{-.2cm}\begin{bmatrix}
0 & 1 & 0 & 0 & 0 \\
-2 & -3 & 0 & 0 & 0 \\
0 & 0 & 0 & 1 & 0 \\
0 & 0 & -1.5 & -2.5 & 0 \\
0 & 0 & 0 & 0 & -4
\end{bmatrix}\hspace{-.1cm},\
A_d\hspace{-.1cm}=\hspace{-.2cm}
\begin{bmatrix}
1 & 0.1 & 0 & 0 & 0 \\
-0.2 & 0.7 & 0 & 0 & 0 \\
0 & 0 & 1 & 0.1 & 0 \\
0 & 0 & -0.15 & 0.75 & 0 \\
0 & 0 & 0 & 0 & 0.6
\end{bmatrix}.
\end{align*}
}

\vspace{-.3cm}
For this example, we obtained {$t^{l}_{c}=0.512 s,t^{s}_{c}=0.581 s$}, respectively, in the CT case, as well as {$t^{l}_{d}=0.374 s,t^{s}_{d}=0.482 s$} for the DT case. All times were slightly higher than the ones for the previous 2-dimensional system, however the difference between the LP and SDP execution times were still very low.
\subsection{Examples 3--6: IEEE 68-Bus, 90-Bus, 150-Bus, and 250-Bus \mo{Systems}}
The third example is an IEEE 68-bus system~[\cite{chatterjee2024grid}] with the following components: 16 generators modeled using swing equations, 5 loads represented as dynamic models, and a network of transmission lines and transformers. The state-space equations are formed from:\\
$\bullet$ Generator Dynamics: $\dot{\omega}_i = \frac{1}{M_i} \left(P_{m,i} - P_{e,i} - D_i \omega_i \right)$, and $\dot{\delta}_i = \omega_i$,
    where the state variables \( \delta_i \) and \( \omega_i \) represent the rotor angle and rotor speed deviation of generator \( i \), respectively. Moreover, the parameters \( M_i \), \( D_i \), \( P_{m,i} \), and \( P_{e,i} \) denote the inertia constant, damping coefficient, mechanical power input, and electrical power input of generator \( i \), respectively;\\
    $\bullet$ Load Dynamics: represented as algebraic or differential equations, and\\
    $\bullet$ Network: power flow equations connecting buses.
    
The system's nonlinear state vector is:
\[
x = [\delta_1 \ \omega_1 \ \delta_2 \ \omega_2  \dots \delta_{16} \ \omega_{16}, V_{\text{Load}} \ \theta_{\text{Load}}]^\top,
\]
where \( \delta_i \) and \( \omega_i \) correspond to the 16 generators, while
\( V_{\text{Load}}\) and \(\theta_{\text{Load}} \) belong to dynamic loads. The linearized matrices were also extracted from~[\cite{chatterjee2024grid}], and the sampling time is $0.01$ s. In the CT case, the LP test took {$t^{l}_{c}=0.532 s$}, while the SDP test took {$t^{s}_{c}=0.831 s$} to complete. Further, in the DT case, we obtained {$t^{l}_{d}=0.381 s$} and {$t^{s}_{d}=0.575 s$}, respectively. As observed, by increasing the number of states the time difference between the LP and SDP methods becomes more considerable.

This fact was better illustrated when we further increased the system dimension, considering IEEE power system case studies with 90, 150, and 250.  as Examples 4, 5, and 6, respectively. The procedure follows a similar approach to Example 3, and the corresponding equations and parameters remain unchanged. The sampling time is also set to $0.01 s$. The execution times for the CT case and for each SDP and LP method were CT: {$t^{s}_{c}=2.902 s$}, {$t^{l}_{c}=0.682 s$}, DT: {$t^{s}_{c}=0.620 s$}, {$t^{l}_{c}=2.154 s$} (for the IEEE 90-Bus system), CT: {$t^{s}_{c}=32.022 s$}, {$t^{l}_{c}=32.022 s$}, DT: {$t^{s}_{d}=5.509 s$}, {$t^{l}_{d}=2.034 s$} (for the IEEE 150-Bus system), and CT: {$t^{s}_{c}=201.483 s$}, {$t^{l}_{c}=22.946 s$}, DT: {$t^{s}_{d}=67.108 s$}, {$t^{l}_{d}=19.451 s$} (for the IEEE 250-Bus system).

\subsection{\sa{Examples 7: 300-Bus Power system}}
\begin{figure}[htbp]
\begin{center}
\includegraphics[width=4.36cm]{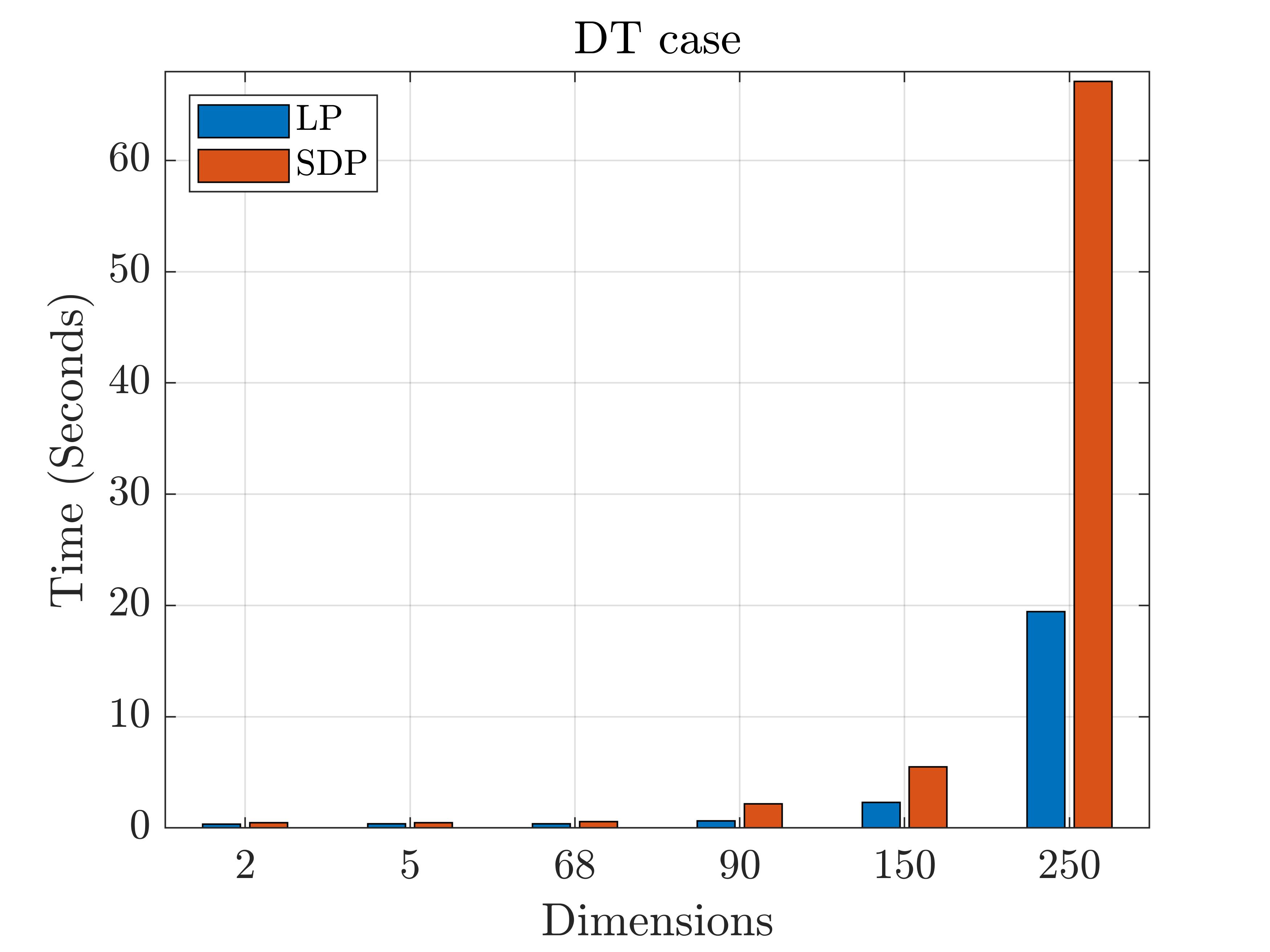}
\includegraphics[width=4.36cm]{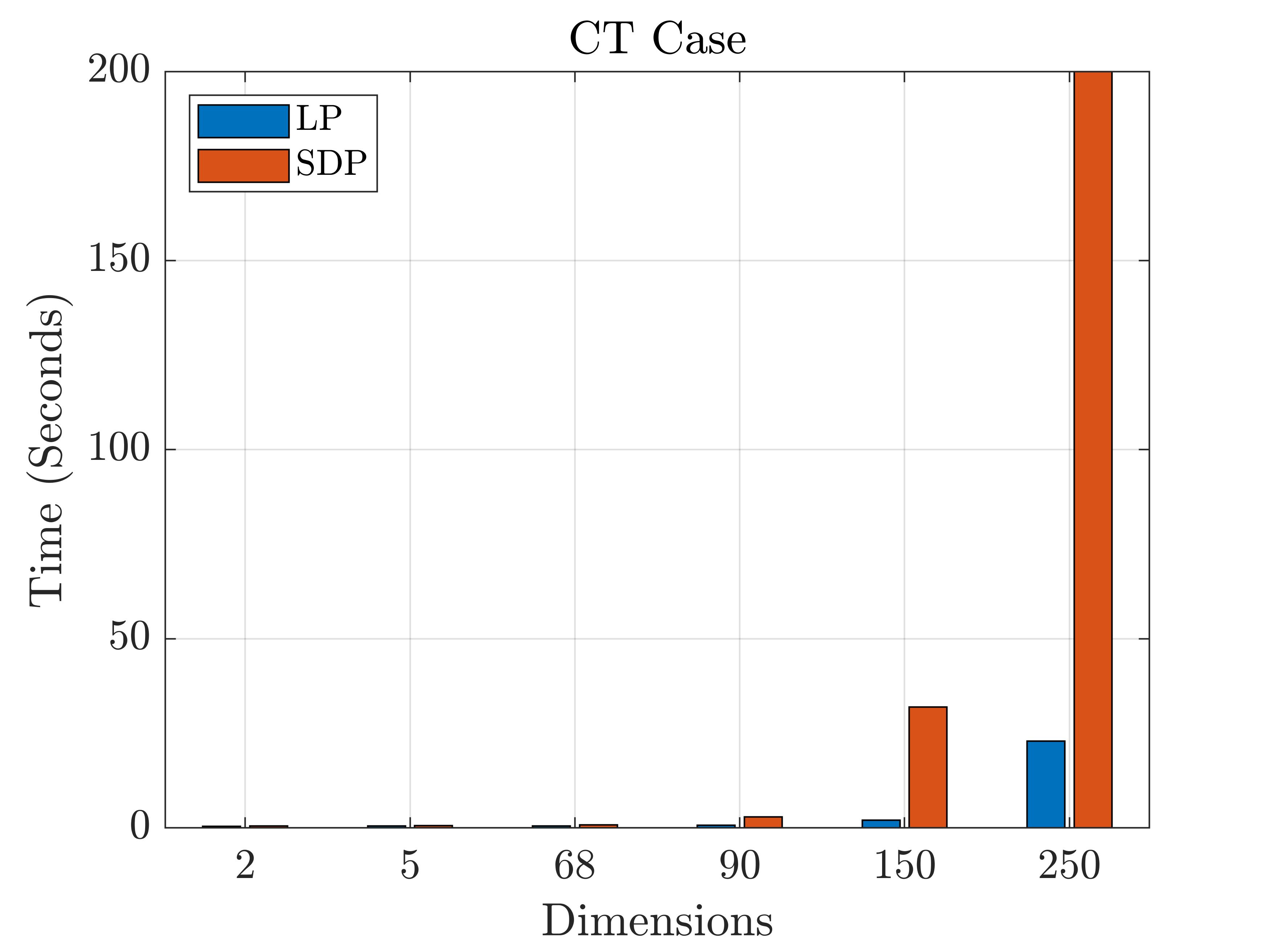} 
\vspace{-.5cm}
\caption{\small The results of applying the SDP and LP methods for the DT case (\mo{left}), and the CT case (\mo{right}). The vertical axis denotes the required computation time, while the horizontal axis shows the dimensions of the systems.\label{fig:DT_time}}
\end{center}
\end{figure}

\sa{In the 7th example, a system with higher dimensions (300-Bus Power system), yet structurally similar to Examples 3–6, is selected~[\cite{vma9-wk20-25}]. It is worth noting that, in this case, the SDP methods for both the DT and CT cases resulted in timeout errors during execution. The execution times for both methods were CT: {$t^{s}_{c}$ : timeout}, {$t^{l}_{c}=26.447 s$}, DT: {$t^{s}_{d}$ : timeout}, {$t^{l}_{d}=24.212 s$}, respectively.}
All the corresponding computation times required to solve the LP and SDP tests to the above examples are shown in Figure \ref{fig:DT_time} for the DT case (top) and the CT case (bottom), respectively. \sa{In Example 7, the SDP method is not included in Figure \ref{fig:DT_time} due to running out of memory and being time-consuming, which resulted in a timeout.} The histograms illustrate the computation times as a function of dimension. Based on the results, for both DT and CT cases, as the dimension increases. The computation time for the SDP method exhibits an almost exponential growth. In contrast, the LP method shows a much slower increase, demonstrating its efficiency in higher dimensions. 
\vspace{-.2cm}
\begin{table}[h!]
\centering
\caption{\small figure methods across 7 case studies for the DT case (top) and the CT case (bottom)}
\label{tab:dt_case}
\begin{tabular}{l c | c c c}
\hspace{1.5cm}{SDP} & & \multicolumn{2}{c}{{LP}} \\
\hline
{Memory (MB)} & {Time (s)} & {Memory (MB)} & {Time (s)} \\
\hline
\hspace{0.9cm}1.008 & 0.463 & 0.847 & 0.353 \\
\hspace{0.9cm}1.847 & 0.482 & 0.952 & 0.374 \\
\hspace{0.9cm}4.021 & 0.575 & 1.844 & 0.381 \\
\hspace{0.9cm}9.326 & 2.154 & 3.518 & 0.620 \\
\hspace{0.85cm}19.655 & 5.509 & 9.225 & 2.034 \\
\hspace{0.75cm}182.146 & 67.108 & 57.351 & 19.451 \\
\hspace{0.75cm}673.224 & timeout & 118.781 & 24.212 
\end{tabular}
\vspace{.2cm}
\begin{tabular}{l c | c c c}
\hspace{1.5cm}{SDP} & & \multicolumn{2}{c}{{LP}} \\
\hline
{Memory (MB)} & {Time (s)} & {Memory (MB)} & {Time (s)} \\
\hline
\hspace{0.9cm}1.186 & 0.540 & 1.149 & 0.448 \\
\hspace{0.9cm}1.457 & 0.581 & 1.187 & 0.512 \\
\hspace{0.9cm}1.735 & 0.831 & 1.274 & 0.532 \\
\hspace{0.9cm}8.655 & 2.902 & 1.654 & 0.682 \\
\hspace{0.85cm}74.946 & 32.022 & 5.119 & 2.099 \\
\hspace{0.75cm}453.015 & 201.483 & 57.582 & 22.946 \\
\hspace{0.65cm}2103.114 & timeout & 125.230 & 26.447 
\end{tabular}
\end{table}
\sa{It is worth mentioning that for low-dimensional systems (dimension $\leq$ 10), the LP method uses approximately 50 to 100 MB, and the SDP method uses approximately 150 to 250 MB. For medium-dimensional systems (dimension $\approx$ 50--100), the LP method uses approximately 120 to 250 MB, and the SDP method uses approximately 800 MB to 1.5 GB. For high-dimensional systems (dimension $\geq$ 200), the LP method uses approximately 300 to 600 MB, while the SDP method uses approximately 2.5 to 5 GB. Additionally, the computation times reported in Table 1 were averaged over five independent runs for each test case to minimize the impact of transient system fluctuations, caching effects, or background processes.}

Finally, the computation times, as well as the corresponding memory usages for all examples are gathered in Table \ref{tab:dt_case}. 
According to this table, in both DT and CT cases, as the dimension and consequently the computation time increases, the required memory also increases. This increase in memory, similar to the trend in computation time, follows an almost exponential pattern for the SDP test. However, in the LP test, this increase does not have a steep increasing trend.
\section{Conclusion and Future Work}
 A novel method was presented in this work for assessing the asymptotic stability of equilibrium points in continuous-time and discrete-time nonlinear autonomous systems. Indirect Lyapunov techniques were employed, and system dynamics were linearized using Jacobian matrices to formulate stability conditions. The approach replaced conventional semi-definite programming methods with computationally efficient linear programming  formulations, significantly reducing the computational complexity, especially for high-dimensional systems. Stability conditions were established through matrix transformations and structural insights into the system, ensuring scalability. Future work will consider providing robust versions of such conditions for bounded-error and uncertain systems, and deriving similar conditions for input-to-state stability.
\balance
\bibliography{ifacconf}             
                                                   







\end{document}